\documentclass{aa}
\usepackage{graphicx}
\begin{document}
%
\title{Deep I-band Imaging of $z$=5.99 Quasar}

   \subtitle{}

   \author{Valentin~D.~Ivanov}

   \institute{European Southern Observatory,
              Ave. Alonso de Cordova 3107, Casilla 19,
              Santiago 19001, Chile\\
              \email{vivanov@eso.org}
             }

   \date{Received ...; accepted ...}

\abstract{
Deep I-band imaging was carried out to search for the optical
counterpart of the X-ray jet candidate near SDSS 1306+0356,
reported by Schwartz (2002, astro-ph/0202190).
The data suggest that the extended X-ray source may be a jet,
related to a galaxy rather than to the quasar itself.

\keywords{quasars: general -- Galaxies:
jets -- Galaxies: high redshift}
}

\maketitle

\section{Introduction}

Synchrotron jets are observed in many quasars and nearby radio
galaxies. They have been detected in a broad range of wavelengths,
from radio to hard X-ray (Scarpa et al. \cite{sca99}, Schwartz
et al. \cite{sch00}, Sambruna et al. \cite{sam02}, and Harris \&
Krawczynski \cite{har02}). The highest redshift quasars are among
the earliest objects in the Universe. The X-ray emission produced
close to the central black hole is related intimately to the
processes in the central engine and therefore offers the
possibility to probe directly the processes occurring in the
quasar nuclei.

Recently, Brandt et al. (\cite{bra02}) carried out a $Chandra$
survey of the quasars with highest known redshifts up to date,
spanning the range from $z$=5.8 to 6.3. All three objects
targeted by this program were detected. Schwartz (\cite{sch02})
found a serendipitous source 23.3 arcsec to the North-East of
the quasar SDSS J130608.26+035626.3 (SDSS 1306+0356 hereafter;
$z$=5.99, Fan et al. \cite{fan01}). The source was extended
over $5\times 2$ arcsec box, aligned with the quasar. Schwartz
suggested that it can be a jet, associated with the quasar.
The goal of this project is to find an optical counterpart of
the X-ray source, and to explore its nature.

\section{Observations and Data Reduction}

\begin{figure}
\resizebox{\hsize}{!}{\includegraphics{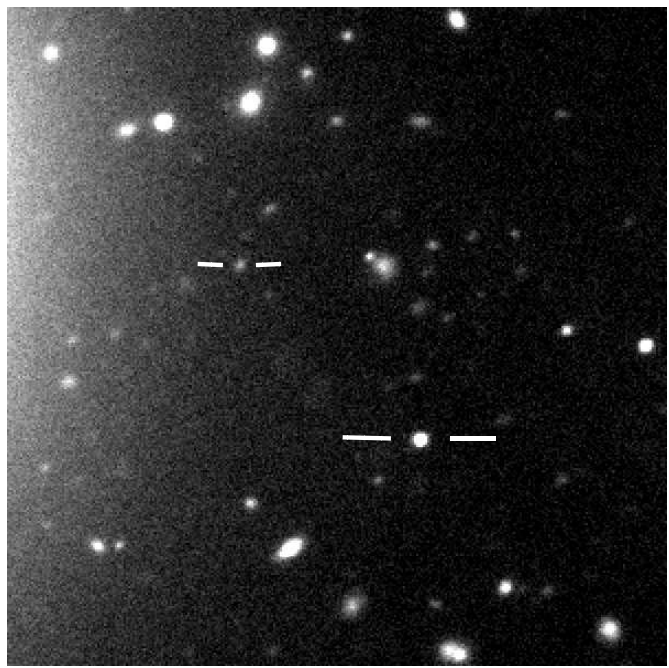}}
\caption{The field of SDSS 1306+0356. The quasar, and the optical
counterpart of the extended $Chandra$ source are indicated,
respectively with long and short dashes. The field is $1\times1$
arcmin. North is up and East is to the left. The higher background
level to the East is caused by a bright nearby star.}
\label{Fig1}
\end{figure}

Deep Bessel I-band imaging was carried out at the VLT with FORS1.
A series of four 787 sec images was obtained on Mar 18, 2002. The
standard data reduction was applied, and the four images were
combined. The seeing measured on multiple unresolved sources is
0.56 arcsec, with a scale of 0.2 arcsec $\rm px^{-1}.$ The outer
parts of the images were affected by reflection from a bright
nearby star but the quasar itself was not contaminated. Gray scale
reproduction of the final image is shown in Figure~\ref{Fig1}

\section{Results}

\begin{figure}
\includegraphics[scale=0.36]{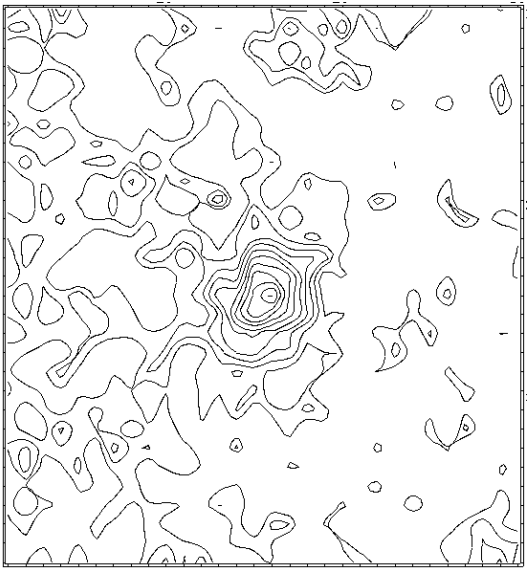}
\caption{Contour plot of the best candidate for an optical
counterpart to the X-ray source. The field of view is 6x6
arcsec, tic marks equal 0.2 arcsec (1 pixel). North is up
and East is to the left. The contours span a range from
23.06 to 26.77 mag $\rm arcsec^{-1},$ and are equally
spaced in flux.}
\label{Fig2}
\end{figure}

\begin{figure}
\includegraphics[scale=0.38]{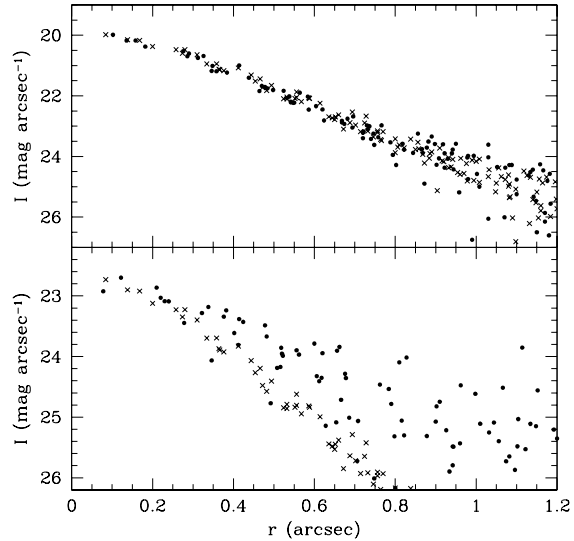}
\caption{Radial profiles of SDSS 1306+0356 (top, solid dots),
and the best candidate for an optical counterpart to the X-ray
source (bottom, solid dots), in arbitrary units. The radial
profiles of a foreground star are shown for comparison (crosses). 
The sky background was subtracted locally for each object.}
\label{Fig3}
\end{figure}

Two candidates for an optical counterpart were discovered near
the location of the X-ray source. The first one is at
RA=13:06:09.34 Dec=+03:56:42.1 (J2000), 22.7 arcsec to the
North-East from the quasar. It is extended in NW-SE direction
(Figure~\ref{Fig2}). The source spans at least 1.5 arcsec along
the major axis, corresponding to a projected diameter of at
least 9 kpc at the redshift of the quasar. We assumed
$\rm H_0=65~km~s^{-1}~Mpc^{-1}, ~\Omega_0=0.3,
~\Omega_\Lambda=0.7, \rm ~and ~q_0=-0.55,$ same as in Schwartz
(\cite{sch02}). The radial profiles
of the quasar and the source are plotted in Figure~\ref{Fig3}.
A bright star is shown for comparison. Clearly, the source is
well resolved, while the quasar is not. The total apparent I-band
magnitudes of the source is $23.01\pm0.11$ mag.

The second candidate is located at RA=13:06:09.19 Dec=+03:56:39.3
(J2000). It is unresolved and has I-band brightness of
$24.81\pm0.20$ mag. It was discarded from further considerations
because its position differs from the location of the X-ray
source by more than 4 arcsec, which exceeds the accuracy of our
coordinates $(\sim0.5$ arcsec, as estimated from the quasar
coordinates).
The apparent I-band magnitudes of the quasar was measured as well:
$20.65\pm0.05$ mag.

The transverse elongation of the optical source with respect to
the elongation of the X-ray source casts a strong doubt on the
interpretation offered by Schwartz (\cite{sch02}). The data
indicates a foreground galaxy, rather than UV/optical emission
from a jet associated with SDSS 1306+0356.

However, Schwartz (\cite{sch02}) demonstrated that the extended
X-ray source is statistically significant. It is comprised of 7
counts, which exceeds by far the expected background flux of
0.13 counts. He estimated that the probability of observing 7
photons in a 10 $\rm arcsec^2$ box is $5.8\times10^{-9}.$

What are the chances of observing a galaxy with the given magnitude
in such a close proximity $(\sim$0.6 arcsec) from the X-ray source?
An integration of the I-band number counts provided by Yasuda
et al. (\cite{yas01}) suggests that $3.8\times10^5$ galaxies
brighter that I$\sim$23 mag can be found per square degree.
Therefore, the probability of random coincidence of the X-ray
source and a galaxy is $\sim0.01.$

An alternative estimate can be obtained by using the surface
density of optically detected objects near the X-ray source.
There are 6 such objects within 10$\times$10 arcsec area. Thus,
the probability is $6\times(0.6/10)^2\sim0.02.$

In the light of the new data, the most likely explanation is
that the X-ray source, if it is a jet, is associated with the
galaxy, rather than the quasar. Multiband observations are
needed to determine the redshift of the host. The bright nearby
star (SAO\,119762, K0, V=8.32 mag) affects the sky background
preventing meaningful galaxy number counts in the vicinity of
the quasar.

\begin{acknowledgements}
The author is grateful to members of Paranal Science Operations
team for carrying out the observations in service mode, and to
the referee Dan Schwartz for the useful comments that helped to
improve this letter.
\end{acknowledgements}

\end{document}